\documentclass[conference]{IEEEtran}
\IEEEoverridecommandlockouts
\usepackage{cite}
\usepackage{amsmath,amssymb,amsfonts}
\usepackage{algorithmic}
\usepackage{caption}
\usepackage{subcaption}
\usepackage{graphicx}
\usepackage{textcomp}
\usepackage{xcolor}
\usepackage{enumitem}
\usepackage{acronym}
\usepackage{url}
\usepackage[binary-units=true, per-mode=symbol]{siunitx}
\def\BibTeX{{\rm B\kern-.05em{\sc i\kern-.025em b}\kern-.08em
    T\kern-.1667em\lower.7ex\hbox{E}\kern-.125emX}}
\begin{document}

\setlength{\textfloatsep}{3pt} 
\setlength{\floatsep}{3pt} 
\setlength{\intextsep}{3pt} 
\setlength{\dbltextfloatsep}{2pt} 
\setlength{\dblfloatsep}{2pt} 
\setlength{\abovecaptionskip}{2pt}
\setlength{\belowcaptionskip}{2pt}


\title{\textsc{Echoes:} a 200 GOPS/W Frequency Domain SoC with FFT Processor and I$^2$S DSP for Flexible Data Acquisition from Microphone Arrays}

 \author{\IEEEauthorblockN{Mattia Sinigaglia\IEEEauthorrefmark{1},
 Luca Bertaccini\IEEEauthorrefmark{2},
 Luca Valente\IEEEauthorrefmark{1},
 Angelo Garofalo\IEEEauthorrefmark{1}\IEEEauthorrefmark{2},
 Simone Benatti\IEEEauthorrefmark{3},
 Luca Benini\IEEEauthorrefmark{1}\IEEEauthorrefmark{2}, \\
 Francesco Conti\IEEEauthorrefmark{1},
 and Davide Rossi\IEEEauthorrefmark{1}}
 
 \IEEEauthorblockA{\IEEEauthorrefmark{1}Department of Electrical, Electronic and Information Engineering (DEI), University of Bologna, Italy}
 \IEEEauthorblockA{\IEEEauthorrefmark{2}Integrated Systems Laboratory (IIS), ETH Z\"{u}rich, Switzerland}   
 \IEEEauthorblockA{\IEEEauthorrefmark{3}Department of Science and Methods for Engineering (DISMI), University of Modena e Reggio Emilia, Italy}}

\maketitle




\begin{abstract}
Emerging applications in the IoT domain require ultra-low-power and high-performance end-nodes to deal with complex near-sensor-data analytics. Domains such as audio, radar, and Structural Health Monitoring require many computations to be performed in the frequency domain rather than in the time domain. 
We present \textsc{Echoes}, a System-On-a-Chip (SoC) composed of a RISC-V core enhanced with fixed- and floating-point digital signal processing (DSP) extensions and a Fast-Fourier Transform (FFT) hardware accelerator targeting emerging frequency-domain application. The proposed SoC features an autonomous I/O engine supporting a wide set of peripherals, including Ultra-Low-Power radars, MEMS, and digital microphones over I$^2$S protocol with full-duplex Time Division Multiplexing DSP mode, making \textsc{Echoes} the first open-source SoC which offers this functionality enabling simultaneous communication with up to 16 I/Os devices. \textsc{Echoes}, fabricated with 65nm CMOS technology, reaches a peak performance of 0.16 GFLOPS and a peak energy efficiency of 9.68 GFLOPS/W on a wide range of floating and fixed-point general-purpose DSP kernels. The FFT accelerator achieves performance up to 10.16 GOPS with an efficiency of 199.8 GOPS/W, improving performance and efficiency by up to 41.1$\times$ and 11.2$\times$, respectively, over its software implementation of this critical task for frequency domain processing.

\end{abstract}


\newcommand{\todo}[1]{\noindent\textit{\color{red}\textbf{}~#1}}

\acrodef{DSP}{Digital Signal Processing}
\acrodef{SHM}{Structural Health Monitoring}
\acrodef{HWPE}{Hardware Processing Engine}
\acrodef{ML}{Machine Learning}
\acrodef{FFT}{Fast-Fourier Transform}
\acrodef{FIR}{Finite Impulse Response}
\acrodef{TDM}{Time Division Multiplexing}
\acrodef{SoC}{System-On-a-Chip}
\acrodef{FPU}{Floating Point Unit}
\acrodef{HWPE}{Hardware Processing Engine}
\acrodef{RTL}{Register Transfer Level}  
\acrodef{SDC}{Synopsys Design Constraint}
\acrodef{IC}{Integrated Circuit}
\acrodef{FMA}{Fused Multiply and Accumulate}

\renewcommand{\baselinestretch}{0.84}\selectfont

\section{Introduction and Related Work}

\begin{figure*}[t!]
\centerline{\includegraphics[width=\textwidth]{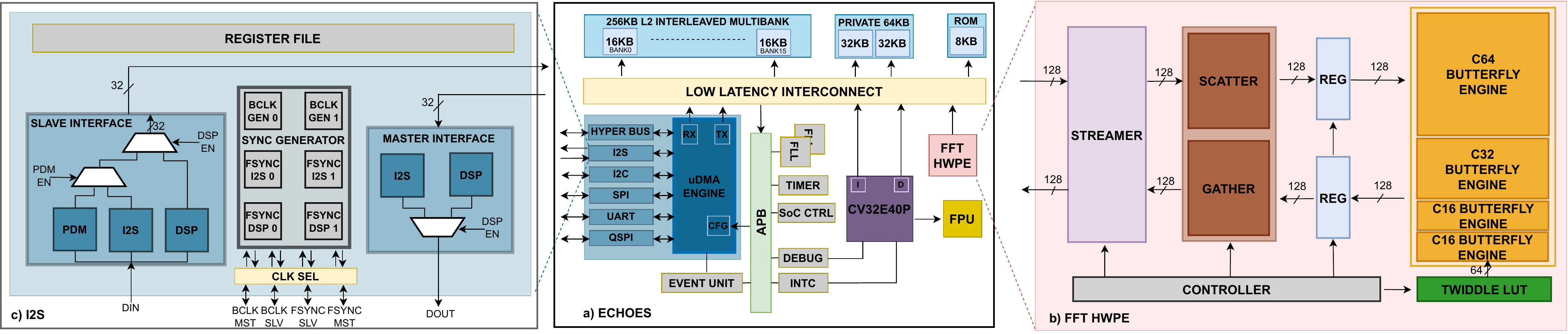}}
\caption{a) Overview of \textsc{Echoes} SoC and details of: b) FFT HWPE accelerator c) I$^2$S with DSP implementation}
\label{fig:echoes_archi}
\vspace{-0.5mm}
\end{figure*}

A wide range of applications in the IoT domain requires powerful end nodes capable of dealing with the increasing complexity of near-sensor-data analytics algorithms. A common feature of these battery-powered systems is the need to operate within a power envelope of less than \SI{100}{\milli\watt}, which, coupled with the high-performance requirements of the algorithms, leads to extremely tight energy efficiency requirements. Several emerging application fields such as audio, gesture recognition, or \ac{SHM} perform computations such as noise canceling, filtering, and equalization in the frequency domain rather than in the time domain, exploiting emerging sensors such as MEMS, radars, and microphone arrays.

Due to the nature of these applications, devices must be powerful but, at the same time, must also have a tiny physical footprint and power consumption. An extreme example is that of hearing aids \cite{KS10}. The minimum power for a given function can be achieved with fully specialized dedicated hardware \cite{Gerlach2021},  but this approach comes with limits in flexibility, and post-fabrication updates \cite{5297793}.
An emerging trend is to couple flexibility with efficiency, fully programmable microcontroller cores enhanced with \ac{DSP}-oriented instruction set extensions. This solution provides enough flexibility to deal with complex and fast-evolving frequency-domain algorithms. In this way, the same device can target various application fields. For example,  structural monitoring analyzing vibrations \cite{endaq}, noise cancellation for consumer applications such as headphones \cite{WH-1000XM5} \cite{700UC} or smart glasses \cite{RBSSG}, gesture recognition, as well as movement and object detection based on radar processing \cite{9381994}\cite{9092991}.
Since these devices must be cheap and produced at a very large scale to be economically viable, this "one-size-fits-many" design approach maximizes the usefulness of emerging \ac{DSP} architectures.




On the other hand, to meet strict application performance requirements, \ac{DSP} devices must run fast and at low power selected key kernels that are 1) highly computationally intensive and 2) used repeatedly. A prime example is that of \ac{FFT}, which plays a key role in \ac{DSP} algorithms, often taking a large portion of the overall computation time. Augmenting a flexible architecture with dedicated \ac{FFT} accelerators is a good strategy to increase performance and efficiency. Most \ac{FFT} accelerators \cite{8240277}\cite{20140171} have been designed with internal memories to host and internally reshuffle butterfly data. However, such internal buffers usually occupy most of the area dedicated to the accelerator, increasing their cost ~\cite{bertaccini2021buffer}. To mitigate this issue, Baas~\cite{748190} proposed an \ac{FFT} accelerator including a small cache, while other implementations share the memory with cores such as \cite{mckeown2010fft}. Furthermore, when addressing a wide range of applications, \ac{FFT} accelerators that support multiple data types are desirable. Using such designs gives a chance to lower the precision, whenever possible, to improve performance and energy efficiency.

Among audio applications, microphone arrays are becoming widespread as many applications require the exploration and characterization of  background sound and/or multiple, spatially distributed noise sources. One way to overcome the limitations of perception of speech intelligibility in time and space in varying noisy environments is the use of microphone arrays, which provides many advantages: directional reception of the sound, spatial localization of the target speaker, or noise suppression of point sources \cite{AMAFDSC}. Beamforming algorithms take advantage of multi-microphone arrays to perform suppression of unwanted contamination sources\cite{BTMASS}. Microphones in digital devices use Inter-\ac{IC} Sound I$^2$S protocol \cite{UM11732}\cite{MS-2275}, which was originally developed for transmitting stereo audio data through a serial line. To perform multi-microphone acquisition, many \ac{SoC}s provide multiple I$^2$S peripherals. This technique leads to high-area occupation and can not be used in small devices, which are often pad-limited. Thus, \ac{SoC}s with strict power and area constraints have to deal with this problem differently. A popular technique is \ac{TDM}. \ac{TDM} allows multiple devices to share the same communication line; each device uses the line for a specific time fraction, and the protocol has to ensure there are no collisions. The limitation of this technique applied to the I$^2$S protocol is the latency introduced by the multiplexing of multiple sources. Many devices follow the DSP mode implementation developed by NXP for implementing low-latency \ac{TDM} \cite{AN3664}.

Some solutions based on RISC-V targeting the audio domain recently appeared on the market. Notable examples are GreenWaves Technologies GAP9 \cite{GAP9} and Telink TLSR9 \cite{TLSR9}. Unfortunately, it is impossible to compare with these commercial \ac{SoC}s since their internal architecture and techniques used to achieve high efficiency are not openly described. From the academic viewpoint, the most recent architectures are those proposed by Nathaniel et al. \cite{rethinking} implementing flexible acceleration through coarse grain reconfigurable arrays (CGRA), and by Jun et al. hosting a hardware accelerator for parallel convolution operation \cite{acoustics4030033} targeting Active Noise Control ANC algorithms. On the other hand, while the proposed architectures are specifically designed to target the audio domain, the proposed \ac{SoC} specifically targets frequency domain applications by coupling a dedicated \ac{FFT} processor with a general-purpose DSP for the flexible acceleration of frequency domain functions.

We present \textsc{Echoes}, a \SI{65}{\nano\meter} \ac{SoC} specialized for frequency-domain \ac{DSP}, centered around an industrially verified extended-ISA RISC-V CV32 core for fixed- and floating-point energy efficient \ac{DSP}. \textsc{Echoes} introduces two main points of novelty. First, it couples the CV32 core with a shared-memory, flexible data-width FFT hardware accelerator to boost conversions to/from the frequency domain, which is a predominant bottleneck in DSP algorithms. Second, it is the first open-source SoC implementing the high configurable full-duplex I$^2$S \ac{TDM} \ac{DSP} mode for communication with up to 16 digital I/Os audio devices. 
\textsc{Echoes} can be used as the central core for \ac{DSP} applications or as an additional device offering \ac{FFT} hardware capabilities. Benchmarked a wide range of floating and fixed point relevant kernels for audio processing, \textsc{Echoes} achieves a peak performance of \SI{0.18}{GFLOPS} and a peak efficiency of \SI{9.68}{GFLOPS/W} when using the CV32 core. The \ac{FFT} accelerator for frequency-domain processing achieves performance up to \SI{10.16}{GOPS} with an efficiency of \SI{199.8}{GOPS/W} which is $11.2 \times$ higher compared to the software implementation and $41.1 \times$ faster. The hardware and software described in this work are open-source, intending to support and boost an innovation ecosystem focusing on ULP computing for the IoT landscape.

\begin{figure*}[t!]
\includegraphics[width=\textwidth ]{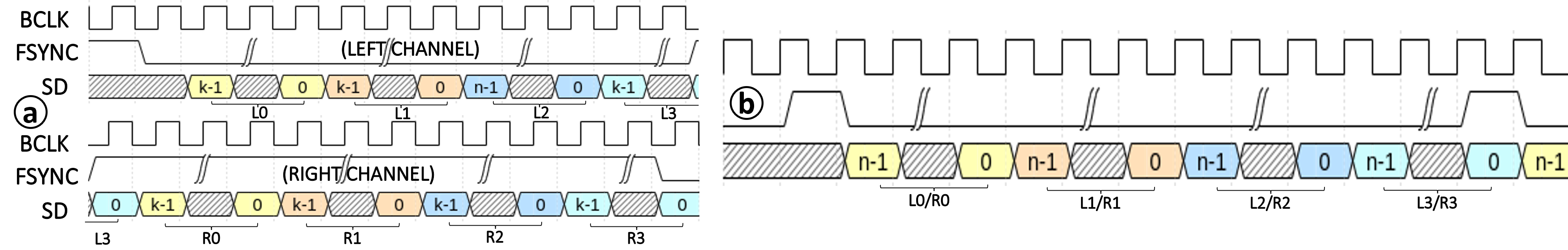}
\caption{Example of I$^2$S Time Division Multiplexing with 4 devices: a) I$^2$S TDM b) I$^2$S TDM DSP Mode. Each color represents a device. The frame length is \textit{n}, while the channel length is $k=n/2$.}
\label{fig:i2s_tdm}
\vspace{-0.5mm}
\end{figure*}


\section{SoC overview}
Fig. \ref{fig:echoes_archi} a) shows the architecture of \textsc{Echoes}. The \ac{SoC} consists of an in-order CV32E40P\cite{CV32E40P} $32$-bit RISC-V core with a $4$-stage pipeline that implements the RV32IMFC instruction set architecture. It features extensions for both floating- and fixed-point energy-efficient \ac{DSP} such as hardware loops, post-increment LD/ST, and Single Instruction Multiple Data (SIMD) such as dot products operating on narrow $16b$ and $8b$ data types \cite{7864441} \cite{8106976}. The core is extended with a \ac{FPU} supporting single-precision floating-point FP32 arithmetic, addition, subtraction, square root, and division, as well as \ac{FMA}, which is a critical kernel for frequency-domain digital signal processing.

The \ac{SoC} is enhanced with an autonomous I/O subsystem coupling an I/O DMA tightly-coupled with a multi-banked system memory~\cite{8106971}. The memory hierarchy of the system is composed of a \SI{256}{kB} L2 memory used to share data among the core, the peripherals, and the accelerator. The L2 memory is organized in $16$ word-level interleaved banks to deliver up to \SI{22.4}GB/S to the master resources. The I/O engine supports a broad set of peripherals such as QSPI, SPI, UART, and I$^2$C that can acquire data from multiple sensors. It also includes an \SI{800}{Mb/s} DDR interface supporting external IoT DRAMs such as Cypress Semiconductor's HyperRAM. Two configurable FLLs provide a dedicated clock source to the processor and peripherals.



The following sections introduce the two architectural contributions of this work relevant for frequency-domain processing and audio applications: the tightly-coupled \textit{FFT Accelerator} and the \textit{I$^2$S \ac{DSP} Time Division Multiplexing}.

\subsection{FFT Accelerator}

To enhance the \ac{DSP} capabilities of \textsc{Echoes}, we integrate a radix-2 decimation-in-time \ac{FFT} \ac{HWPE} into the SoC. \acp{HWPE} are a class of hardware accelerators that share the memory with one or multiple compute cores, thus enabling efficient cooperation between the general-purpose and the domain-specific part of the architecture. Such an approach is uncommon in traditional \ac{FFT} accelerators, which are often decoupled from the core, requiring data copies and large internal buffers. 

Since the butterfly organization of the \ac{FFT} algorithm easily leads to systematic banking conflicts, the \ac{HWPE} implements the scheme recently proposed by \cite{bertaccini2021buffer} to reorder the butterfly sequence and the memory accesses. This scheme requires access to at least $16$ distinct memory banks. It allows the accelerator to load and store data at consecutive memory locations by moving sets of right or left butterfly wings. The samples are then reordered before and after the butterfly computation using internal registers. In this way, banking conflicts can only appear during the final bit-reversed reordering required by the \ac{FFT} algorithm; they are handled by stalling the accelerator pipeline for a single cycle.

The \ac{FFT} accelerator works with fixed-point complex data whose real and imaginary parts are represented with $32/16/8$ bits. We will refer to these data types as $C64/C32/C16$. The maximum number of \ac{FFT} points supported is $512/1024/2048$, respectively for $C64/C32/C16$. The accelerator is programmed by the core, responsible for providing the number of \ac{FFT} points, the data type, the memory address at which the input samples are stored, and the start command. The \ac{HWPE} contains eight $32$-bit memory ports, of which four are used as input and four as output ports, and it is organized in submodules as shown in Fig. \ref{fig:echoes_archi} c). 

The Butterfly Unit is composed of one $C64$, one $C32$, and two $C16$ butterfly engines. As wider butterfly engines can be reused for lower-precision computations, the Butterfly Unit can compute up to $1/2/4$ $C64/C32/C16$ butterflies per cycle. Two sets of four $C64$ Butterfly Registers are located around the Butterfly Unit to reorder inputs and outputs, while the twiddle factors are stored in a lookup table.


\subsection{I$^2$S DSP Time Division Multiplexing}
%

I$^2$S is a standard for digital audio acquisition that defines three main signals: BCKL represents the clock signal; SD (Serial Data) transports the audio data; FSYNC (Frame Synch) selects the stereo channel of transmitted data (Left or Right). The peripheral can act both as a master and a slave.
%
The I$^2$S protocol specifies that the master serially drives data over the SD lines on the falling edges of the BCLK, and the slave samples them on the rising edge. The FSYNC signal discriminates between the Left and Right channels.

In the \ac{TDM} I$^2$S implementation, the multiplexed devices completely bind the time needed to receive both channels from a specific source. Each device of the array first sends the L channel and then the R channel, following the connection order of the array. This order implies that the number of devices introduces significant latency in communication -- a very undesirable effect, particularly for audio processing, which is highly latency sensitive. Fig. \ref{fig:i2s_tdm} a) shows an example of the I$^2$S \ac{TDM} implementation with \SI{4} devices. Considering that each frame is composed of $nBits$ and the number of the multiplexed devices is $K$, each frame of a source has a latency of $\frac{nBits}{2} \times (K+1) \times Tclk$.


To reduce audio signal acquisition latency, \textsc{Echoes} implements a more specialized I$^2$S protocol dealing with the parallel acquisition from multiple microphones: \ac{TDM} \ac{DSP} mode. Devices supporting \ac{DSP} mode  transmit L/R channels one after the other immediately when the FSYNC is asserted. Fig. \ref{fig:i2s_tdm} b) shows the implementation of the \ac{DSP} \ac{TDM} mode with an example of $4$ devices. In this implementation, each device occupies a specific slot during which it sends the entire frame, and the latency to complete the transmission is reduced to a fixed $nBits\times Tclk$.

Fig. \ref{fig:echoes_archi} c) shows the block diagram of the I$^2$S peripheral. It is composed of two interfaces for transmitting and receiving audio data. Each interface can be programmed to act as \ac{DSP} \ac{TDM} or standard I$^2$S. Both have dedicated I$^2$S buses, which allow for independent and full-duplex communications. The core programs the memory-mapped $32$-bit register file, allowing each interface's independent configuration flexibility. By scaling the peripheral clock, the peripheral features $2$ clock dividers for generating the interfaces' BCLK signals. The four dedicated and independent FSYNC blocks generate the FSYNC signal accordingly to the I$^2$S and \ac{DSP} modes for each interface.

This operation mode is designed for multiplexing $16$ devices that work with a maximum \SI{48}{\kilo\hertz} sample rate. The BCLK frequency of the \ac{TDM} \ac{DSP} mode depends on the combination of $nDevices \times frameBits \times sampleRate$. A \ac{SoC} frequency of around \SI{25}{\mega\hertz} (much lower than the maximum achievable in \textsc{Echoes}) is sufficient to support the best available rate when samples have a frame length of $32$ bits. Thanks to the high frequency reached by \textsc{Echoes}, it is possible to read and compute data before $nDevices$ new data have been received and stored. To deal with different implementations of the \ac{DSP} mode, \textsc{Echoes} can also be programmed to choose the clock polarity and program the frame alignment.


\section{Physical Implementation and measurements}



\begin{figure}[t!]
\centerline{\includegraphics[width=0.85\linewidth]{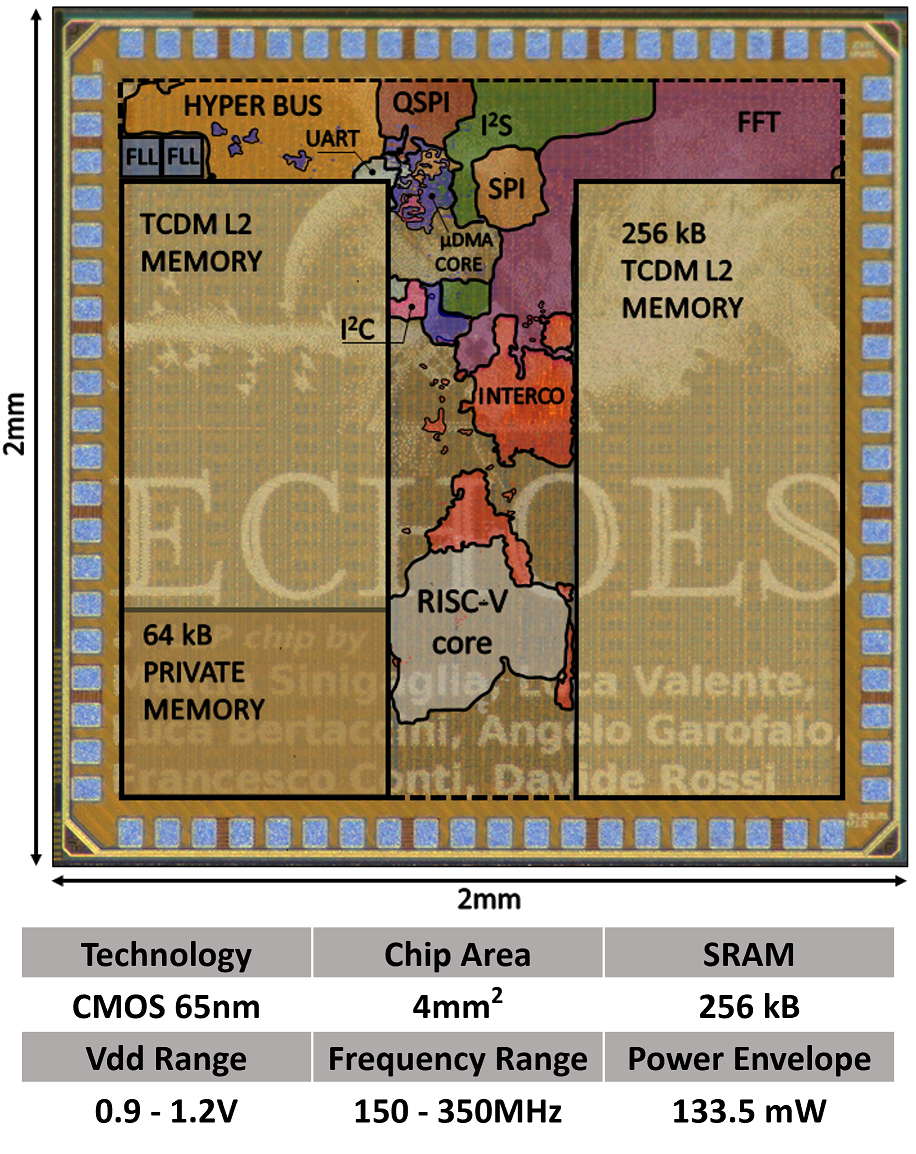}}
\caption{Chip micrograph and specifications}
\label{fig:echos_layout}
\vspace{-0.5mm}
\end{figure}

\textsc{Echoes} has been implemented in TSMC \SI{65}{\nano\meter} technology performing synthesis with Synopsys Design Compiler\cite{synopsis}, Place \& Route with Cadence Innovus\cite{innovus}, and verification and sign-off with Calibre \cite{calibre}.
The die area is \SI{4}{\milli\meter\squared}, partitioned as shown in Fig. \ref{fig:echos_layout} where the majority of the area is occupied by \SI{256}{kB} L2 Memory and the \SI{64}{kB} private memory. The \ac{FFT} accelerator occupies $6\%$, the I$^2$S occupies $2\%$ of the total area, and the core occupies $3\%$.

\begin{figure}[t!]
\centerline{\includegraphics[width=0.85\linewidth]{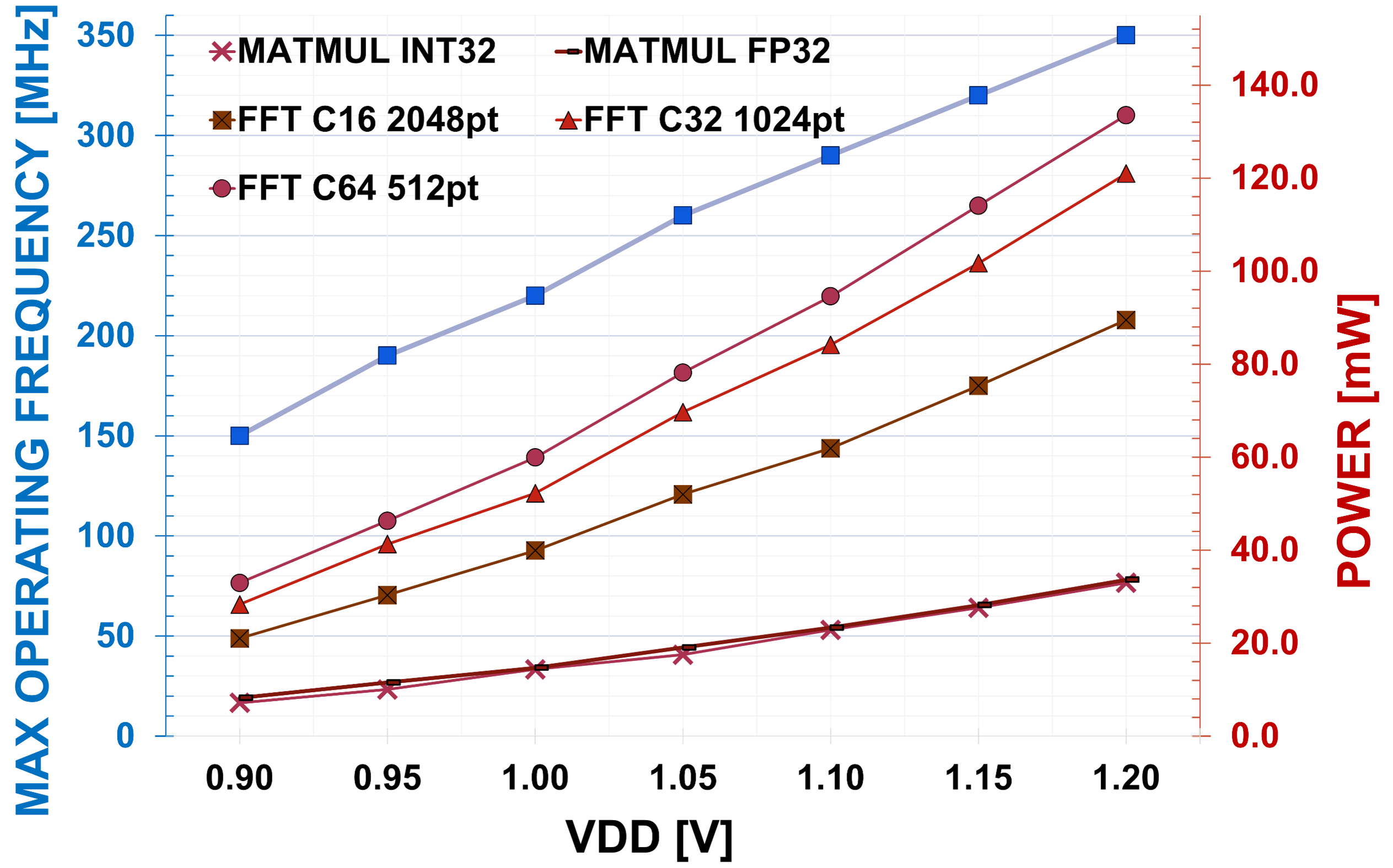}}
\caption{Voltage sweep vs. max frequency vs. power consumption.}
\label{fig:echoes_freq_power}
\vspace{-0.5mm}
\end{figure}

The \textsc{Echoes} chip has been tested and characterized using an Advantest SoC hp9300 integrated circuit testing device. The measurements are taken by executing Matmul and both software and hardware-accelerated FFTs on the chip.
Fig. \ref{fig:echoes_freq_power} reports the maximum operating frequency and the power consumption over \SI{0.9}{\volt} to \SI{1.2}{\volt} for representative of many compute-intensive workloads in the field of audio processing. The Matmuls kernels have been performed using $16\times16$ matrixes and consume respectively \SI{33}{\milli\watt} at \SI{1.2}{\volt} for INT32, and \SI{33.6}{\milli\watt} at \SI{1.2}{\volt} for FP32. The power consumption of the \ac{FFT} accelerator has been measured executing $C64/C32/C16$ workloads respectively on $512/1024/2048$ \ac{FFT} points. The peak power consumption of the \ac{FFT} reaches \SI{133.5}{\milli\watt} at \SI{1.2}{\volt} when executing the FFT $C64$ on $512pt$.

\section{Benchmarking}

\begin{figure}[t!]
\centerline{\includegraphics[width=0.9\linewidth]{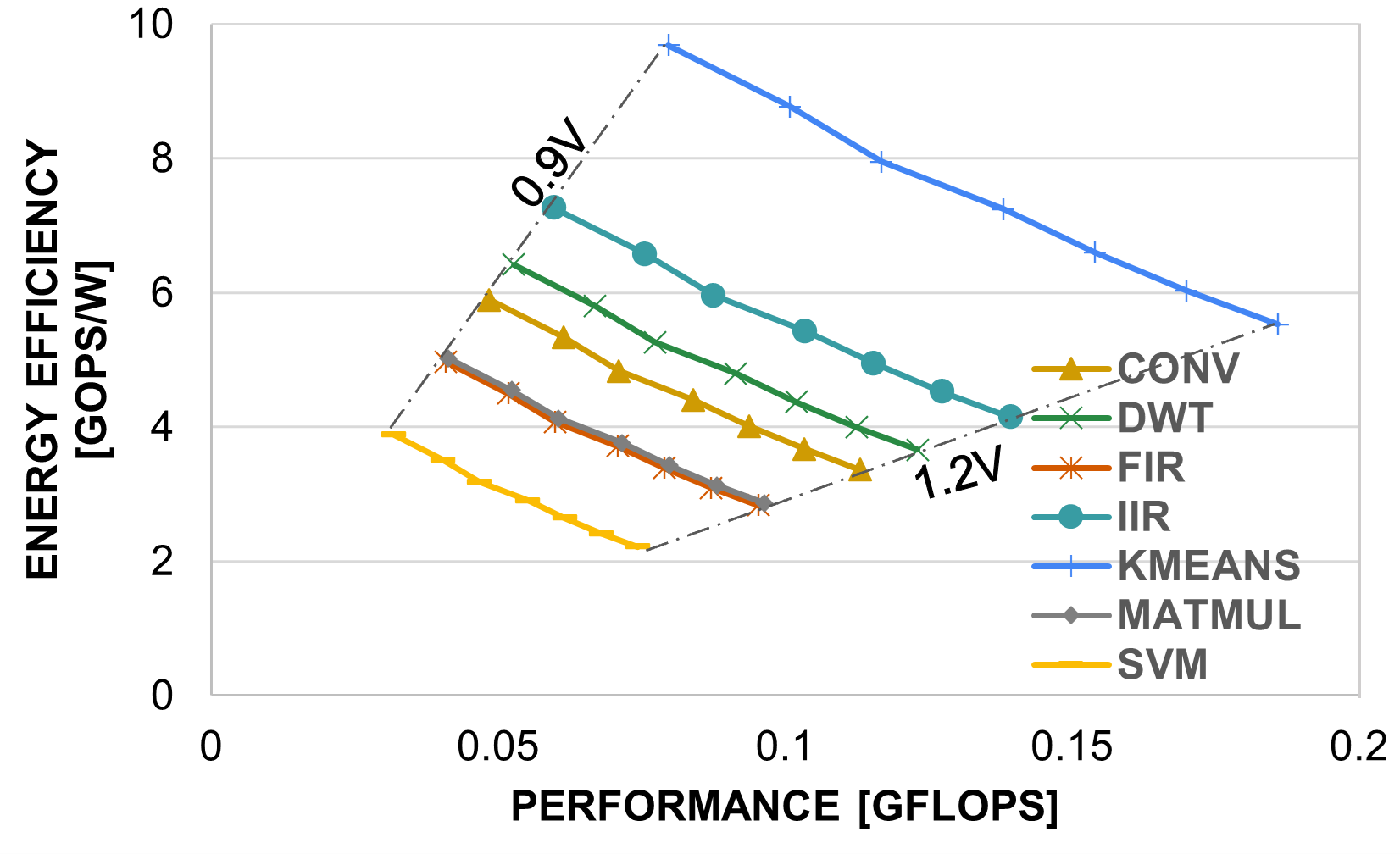}}
\caption{Performance and energy efficiency of DSP kernels on \textsc{Echoes}.}
\label{fig:echoes_energy_eff_ml}
\vspace{-0.5mm}
\end{figure}
\begin{figure}[t!]
\centerline{\includegraphics[width=\linewidth]{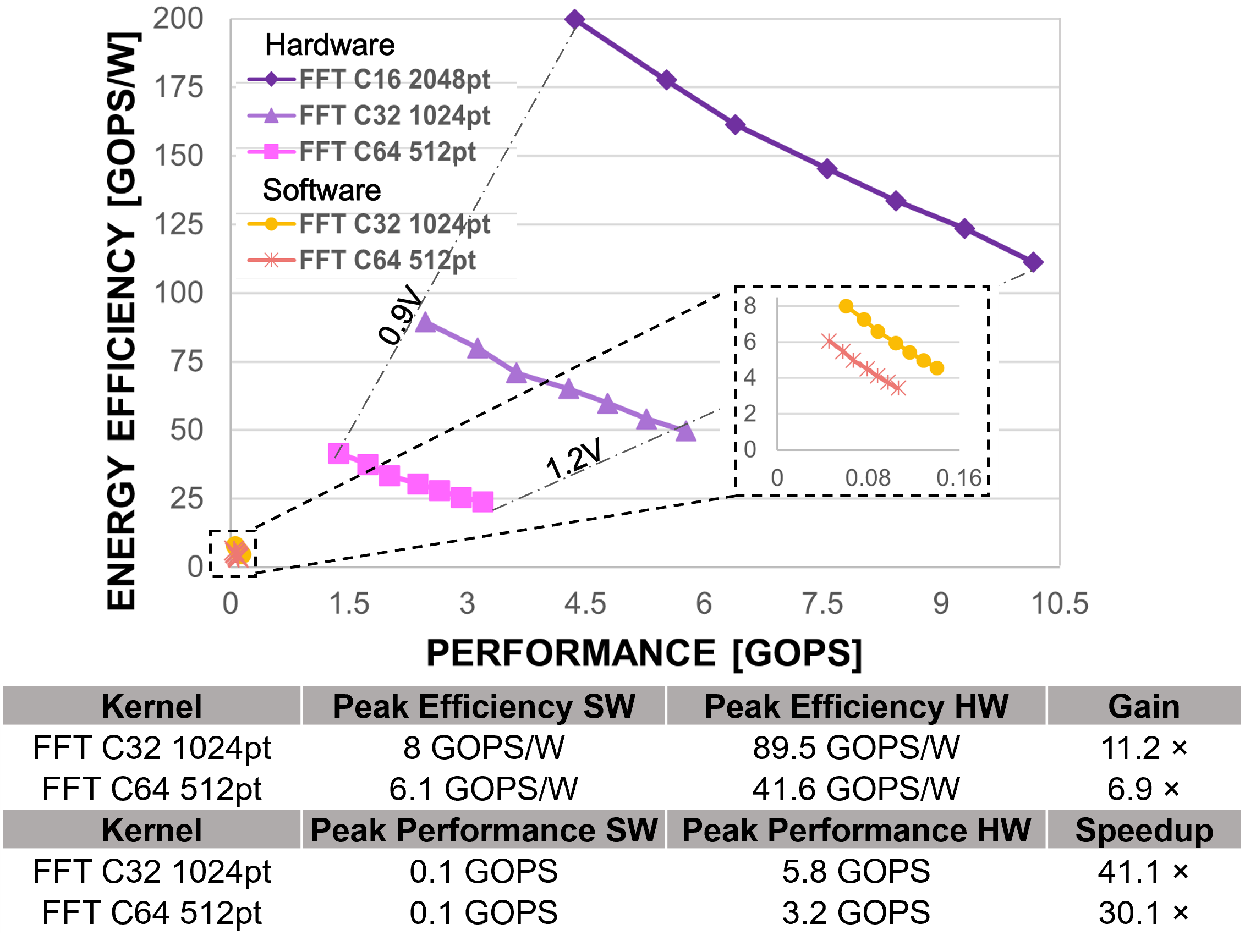}}
\caption{Performance and Energy Efficiency of FFT kernels on \textsc{Echoes}. Speedup and energy gain of FFT computation on the hardware accelerator.}
\label{fig:echoes_energy_eff_dsp}
\vspace{-0.5mm}
\end{figure}

To demonstrate the capabilities and flexibility of \textsc{Echoes}, we used benchmark representative of typical frequency-domain and audio \ac{DSP} workloads.
Fig. \ref{fig:echoes_energy_eff_ml} shows the performance and energy efficiency of the RISC-V core computing FP32 \ac{DSP} kernels. Thanks to the high versatility of the programmable core, \textsc{Echoes} can execute a variety of workloads from frequency- and time-domain processing applications. Fig. \ref{fig:echoes_energy_eff_dsp} reports the performance and the energy efficiency of \ac{FFT} kernels computed using the accelerator and the core, as well as the related speed-up. Thanks to the multiple data sizes supported by the \ac{FFT} accelerator, the precision of a specific kernel can be tuned to the application requirements, enhancing the performance and energy efficiency by $2\times$ when moving from $C64$ to $C32$ formats.


To put our results in perspective, we benchmark the Mel-Frequency Cepstral Coefficients (MFCC) feature extractor and compare it against the solution proposed in \cite{9458491}. Contrarily to \cite{9458491}, where the execution is offloaded to a parallel $8$ cores cluster, in \textsc{Echoes}, we exploit the tightly-coupled cooperation between the single general-purpose core and the specialized \ac{FFT} accelerator, leading to a much more compact form factor of the system ($4mm^2$ vs $10mm^2$).
Despite the fact that the inference of a single MFCC on \textsc{Echoes} is executed in $120 \mu$s, $1.3\times$ slower than \cite{9458491}, due to non-FFT kernels executed on a single core instead of a multi-core cluster, our solution outperforms \cite{9458491} by $2\times$ in area efficiency, despite the less scaled technology node used for implementation.


\section{Conclusion}
We presented \textsc{Echoes}, a \ac{SoC} composed of a RISC-V core enhanced with fixed- and floating-point \ac{DSP} extensions and an \ac{FFT} hardware accelerator targeting emerging frequency-domain applications.
The proposed SoC features an autonomous I/O engine 
able to interface to a wide range of sensors such as MEMS, ULP radars, and digital microphones over I$^2$S protocol with full-duplex \ac{TDM} \ac{DSP} enabling simultaneous communication with up to $16$ I/Os devices. Thanks to the many configurations implemented, \textsc{Echoes} reaches extreme flexibility and makes it the first open-source\footnote[1]{https://github.com/pulp-platform} SoC, which implements all the \ac{TDM} \ac{DSP} configurations available in the market.
The proposed \ac{SoC}, fabricated in TSMC \SI{65}{\nano\meter} technology, can achieve FP32 peak performance up to \SI{0.18}{GFLOPS} at \SI{1.2}{\volt} with an efficiency of \SI{9.68}{GFLOPS/W} at \SI{0.9}{\volt}. The \ac{FFT} accelerator for frequency domain \ac{DSP} processing achieves performance up to \SI{10.16}{GOPS} at \SI{1.2}{\volt} with an efficiency of \SI{199.8}{GOPS/W} at \SI{0.9}{\volt} which is $11.2 \times$ higher compared to the software implementation and $41.1 \times$ faster.
\section*{Acknowledgement}
Supported in part through funding from the ECSEL Joint Undertaking (JU) under grant agreement No 877056 (FRACTAL).
The JU receives support from the European Union’s Horizon 2020 research and innovation programme and Spain, Italy, Austria, Germany, Finland, Switzerland.

\clearpage

\renewcommand{\baselinestretch}{1}\selectfont

\bibliographystyle{IEEEtran}

\bibliography{echoes_bibliography.bib}

\end{document}